\documentclass[a4paper]{jpconf}
\usepackage{graphicx}
\usepackage{cite}
\usepackage{xcolor}
\usepackage{amsmath,amssymb,amsfonts,latexsym,cancel}
\usepackage[normalem]{ulem}

\newcommand{\be}{\begin{equation}}
\newcommand{\ee}{\end{equation}}
\newcommand{\bea}{\begin{eqnarray}}
\newcommand{\eea}{\end{eqnarray}}

\begin{document}
\title{Hadamard and boundary conditions for the Big Bang quantum vacuum}

\author{Pau Beltr\'an-Palau$^{1}$, Sergi Nadal-Gisbert$^{1}$, Jos\'e Navarro-Salas$^{1}$ and Silvia Pla$^{2}$}
\address{${}^{1}$Departamento de Fisica Teorica and IFIC, Centro Mixto Universidad de Valencia-CSIC. Facultad de Fisica, Universidad de Valencia, Burjassot-46100, Valencia, Spain.}
\address{${}^{2}$Theoretical Particle Physics and Cosmology, King’s College London, WC2R 2LS London, United Kingdom.}

\ead{pau.beltran@uv.es, sergi.nadal@uv.es, jnavarro@ific.uv.es \textrm{and} silvia.pla\_garcia@kcl.ac.uk;}

%

%
%\ead{sergi.nadal@uv.es}
%\address{Departamento de Fisica Teorica and IFIC, Centro Mixto Universidad de Valencia-CSIC. Facultad de Fisica, Universidad de Valencia, Burjassot-46100, Valencia, Spain.}

%
%\author{Jos\'e Navarro-Salas}\ead{jnavarro@ific.uv.es}
%\address{Departamento de Fisica Teorica and IFIC, Centro Mixto Universidad de Valencia-CSIC. Facultad de Fisica, Universidad de Valencia, Burjassot-46100, Valencia, Spain.}
%
%\author{and Silvia Pla}

\begin{abstract}
General relativity predicts final-type singularities inside black holes, as well as a cosmological initial-type singularity. Cosmic censorship protects external observers from black hole singularities, while Penrose's Weyl curvature hypothesis protects the smoothness of the initial (Big Bang) singularity. We discuss a simple realization of the Weyl curvature hypothesis by assuming a very early radiation-dominated universe and analytically extending the expansion factor to negative values of conformal time. We  impose time-reversal conditions at the Big Bang to characterize a natural set of preferred vacuum states for quantized matter fields. We implement the prescription of States of Low Energy constructed around the Big Bang obtaining Hadamard states. We also explore the physical implications of these vacua for cosmological dark matter production.

\end{abstract}

\section{Introduction}

The Weyl curvature hypothesis introduced by Penrose  \cite{Penrose79} is a natural and well-motivated way to smooth the Big Bang initial singularity. It can be regarded as complementary to Penrose's cosmic censorship hypothesis \cite{Penrose69},  which protects the outside world from  singularities that are predicted inside black holes. %in the future (inside black holes).

The simplest way to enforce the Weyl curvature hypothesis, and to do so in a way that is consistent with observations, is through a Friedmann-Lema\^{i}tre-Robertson-Walker (FLRW) geometry for spacetime \cite{Newman93}. Furthermore, as we go back in time the expansion factor $a(\tau) $ behaves in a very simple way $a(\tau) \propto \tau$, with respect to the conformal time $\tau$. This makes also natural to analytically extend to negative values the conformal time parameter \cite{Audretsch78}, somewhat mimicking a bouncing geometry at the Big Bang.  The advantage of this picture is that one can easily construct a well-defined vacuum state $|0_{-}\rangle$ at $\tau \to -\infty$. This vacuum state is then perceived as a collection of real particles at $\tau \to +\infty$. The absence of real particles at $\tau \to +\infty $ defines the vacuum $|0_{+}\rangle$. This is an example of gravitation particle creation, as originally established in \cite{Parker66, Parker68, Parker69}. Here we want to emphasize instead the emergent  symmetry of the metric 
\be ds^2 \propto \tau^2 (d\tau^2 - d\vec x^2) \label{eq:FLRW}
\, ,\ee
under the transformation
\be T: \tau \to -\tau \ . \ee
This time-reversal symmetry has been pushed forward in \cite{Large,LetterCPT}. The vacuum state proposed in these references is the $T$-invariant linear combination $|0_{T}\rangle\propto|0_{+}\rangle + |0_{-}\rangle$. The corresponding particle production spectra at late times for both $|0_-\rangle$ and $|0_T\rangle$ is the same for large momenta, a decaying exponential, essentially describing  thermal radiation. 

In this work we analyze the issue of how to construct a vacuum state around the Big Bang $\tau \sim 0$. We note that, the selection of a preferred vacuum in the above-mentioned approaches is based on the asymptotic behaviour at $\tau \sim -\infty$, or  $\tau \sim \pm\infty$. These vacua trivially satisfy the so-called Hadamard condition since they are constructed in the asymptotic regimes of the scale factor, for which the Hubble rate tends to zero.  We recall that an admissible physical state in curved spacetime should have a short-distance behaviour similar to the vacuum in Minkowski space. We refer to this non-trivial property the Hadamard condition \cite{Wald94}. In an expanding universe this requirement was first established in momentum space and referred  as  the adiabatic condition \cite{parker-toms}. In contrast, the construction of sensible vacuum states around $\tau \sim 0$, is more involved.

%It requires to satisfy the so-called Hadamard condition, which is then highly non-trivial.

Selecting initial data to define the vacuum in a rapidly changing region of the spacetime, as is the case for the Big Bang, does not usually render a physically admissible quantum state in the ultraviolet. However, this non-trivial issue can be satisfactorily sorted out by using the states of low energy prescription \cite{Olbermann07,Nadal23}. %}, as we previously showed in detail in \cite{Nadal23}. 
In this approach, the vacuum is defined as the state that  minimizes the energy density smeared over a temporal window. The resulting state depends on the specific window function used and has been checked to satisfy the so-called Hadamard condition.

However, obtaining analytical expressions for the vacuum solutions %that can be analytically parametrized, 
that are also Hadamard is highly complicated around the Big Bang, even for states of low energy. Nevertheless, one can construct approximate Hadamard states that renders a finite  $\langle T_{ab} \rangle$ after renormalization, and can be used to study particle creation and backreaction effects in semiclassical gravity. In this context, we analyze the possible approximated Hadamard vacuum states that can be defined around the Big Bang. %{\color{red} Estudiar el cas $\theta_k=0$ com al que en $\tau=0$ minimitza la densitat d'energia per a $\xi =\frac16$. Renormalitzacio de $\langle T_{ab} \rangle$ i particle creation decay. Despres comentar els SLE aproximats que són més convergents però depenen de la smearing.}

We consider two main applications of the above formalism in the context of gravitation particle production. One is the possible creation of gravitons in a radiation-dominated spacetime. If the vacuum is assumed to be constructed from the conformal modes, the predicted particle creation is exactly zero. This is the usual viewpoint. However, if the vacuum is a state of low energy at the Big Bang a non-trivial creation of (primordial) gravitons is allowed. Therefore, the inflationary paradigm is not the unique scenario to create gravitons. A second application, motivated here by conformal invariance, is the estimation of the mass for a right-handed sterile neutrino as a dark matter candidate. The sterile neutrino is assumed to be decoupled from the other particles of the Standard Model and the production mechanism is gravitational particle creation.

 %We will study the decaying behaviour to large momentum of the number of created particles at late time. Finally, adopting this cosmic view, we will discuss two major consequences of the particle creation phenomena, the estimation of a mass for a right handed sterile neutrino as a dark matter candidate and the prediction of graviton production in a radiation-dominated spacetime.

\section{CPT-invariant vacuum states for scalar fields at the Big Bang: the Hadamard condition} \label{sec:scalars-CPT}

In this section, we introduce the problem under consideration by considering a massive scalar field $\phi$ propagating in a flat FLRW spacetime \eqref{eq:FLRW} with a scale factor of the form 
$a(\tau) \propto \tau$, which corresponds to radiation-dominated universe.  The coupling to the curvature $\xi R\phi^2$ is assumed to be generic. 
It is convenient to expand the quantized field in usual Fourier modes 
\be \label{modecomp}
    \phi(\tau,\vec x)= \int \frac{d^3 k}{\sqrt{2(2\pi)^{3}}}\left(A_{\vec k} e^{i\vec k \vec x} \phi_k(\tau) + A_{\vec k}^{\dagger} e^{-i\vec k \vec x} \phi_k^{*}(\tau)\right) \ ,
\ee
where the creation and annihilation operators satisfy the usual commutation relations.
%($[A_{\vec k}, A^\dagger_{\vec k'}] = \delta^3(\vec k - \vec k')$, etc) with the appropiate normalization condition. %The normalization of the modes is fixed by the condition
%\be \phi_k\phi'{}^{*}_k - \phi_k' \phi_k^* = \frac{2i}{a^2} \ , \ee
%where the prime denotes derivative with respect to the conformal time. 
For our purposes it is convenient  to work with the rescaled Weyl modes %$\varphi \equiv a \phi$ and therefothe rescaled  modes 
$\varphi_k (\tau)\equiv a(\tau) \phi_k(\tau)$. The field equation implies  ($m a=\gamma \tau$)
\be\label{eqm}
  \varphi_k'' + (k^2+\gamma^2\tau^2 ) \varphi_k=0 \ , 
\ee
and normalization condition 
$\varphi_k  \varphi'{}^{*}_k - \varphi_k' \varphi_k^* = 2i  $.
The general solution of \eqref{eqm} can be expressed in terms of parabolic cylindrical functions $D_{\nu}(z)$ \cite{booksf}
\be\label{sol-scalars}
\varphi_k(\tau)= \frac{1}{(2 \gamma)^{1/4}}\Big[C_{k,1}\,  D_{-\frac{1}{2}-2i\kappa }\big(e^{i\tfrac{\pi}{4}}
  \sqrt {2\gamma } \,\tau\big)+C_{k,2}\,  D_{-\frac{1}{2}+2i\kappa}\big(e^{i\frac{3\pi}{4}} \sqrt {2\gamma}\,\tau\big)\Big]\, ,
\ee
where $\kappa=\frac{k^2}{4 \gamma}$. Any choice of $k$-functions $C_{k,1}$ and $C_{k,2}$ defines a set of modes characterizing a given vacuum state. %\footnote{From the Wronskian condition \eqref{Ncondition2} we get the following normalization condition for $C_1$ and $C_2$: $$ \frac{e^{ \pi \kappa}}{2}(|C_1|^2+|C_2|^2)+\frac{\sqrt{2}\cosh{(2 \pi \kappa)}}{\sqrt{\pi}}\textrm{Re}[e^{-i\frac{\pi}{4}}C_1C^*_2\Gamma(\tfrac{1}{2}-2i\kappa)]=1$$} 
%All these vacua are, by construction, invariant under spatial translations, rotations, parity and charge conjugation (which is here trivial since our scalar field is real). Time translation is not a symmetry for an expanding universe, and time-reversal has not been considered so far.  
%As remarked in the introduction, there is no natural way to select a preferred Fock vacua.
%In the radiation-dominated universe, and due to the very special form of the expansion factor in conformal time, 

One can  reduce the freedom in choosing a vacuum by exploiting the time-reversal symmetry  $T: \tau \to -\tau$ of the background $ds^2 \propto \tau^2(d\tau^2 -d\vec x^2$). Since our scalar field theory  is trivially invariant under parity ($P$) and charge conjugation ($C$) we can impose, more generally,  CPT invariance. This means $CPT \varphi(x) (CPT)^{-1}= \varphi(-x)^{\dagger}$.  
%Generically, the action of charge conjugation $C$, parity $P$, and time-reversal $T$ on a classical  scalar field is given by \cite{AGbook} $C:\phi(\tau,\vec{x}) \to \xi_c^*\phi^{*}(\tau,\vec{x}) $; $P:\phi(\tau, \vec x)\to \xi_p^*\phi (\tau, -\vec x)$; $T\phi(\tau, \vec x) \to \xi_t^* \phi^*(-\tau, \vec x)$. [The $\xi$'s are the associated phases of the $C$, $P$, $T$ transformations]. In the quantized theory, $C$ and $P$ are representated as unitary operators, while $T$ is converted into a antiunitary operator. In our case $C$ and $P$ are trivially implemented in the assumed Fourier expansion. Enforcing $T$ is the key ingredient. However,  it seems more useful \cite{LetterCPT, Large} to consider $CPT$ all at once in the analysis (furthermore, we also choose $\xi_c\xi_p\xi_t=1$ \cite{Weinberg} and$CPT \phi(x) (CPT)^{-1} = \phi(-x)^{\dagger}$).
The condition for a CPT-invariant vacuum state takes a very  simple form on the time-dependent part of  scalar field modes
\be \varphi_k(-\tau) =  \varphi_k^*(\tau) \ . \ee 
Using standard properties of the parabolic cylindrical functions it can be easily shown that in terms of the general solution \eqref{sol-scalars}, the condition above implies $C_{k,1}=C_{k,2}^*$.  For our purposes it is convenient to characterize the CPT-invariant condition at $\tau=0$.  We easily find
\be \varphi_k (0)= \varphi_k^{*}(0)\, ,  \qquad
\varphi_{k}' (0) = -\varphi_k^{*\prime}(0)\, . \ee 
The general solution with the above constraints and the normalization condition  can be  parametrized  as 
\be \label{cptscalar0}\varphi_k(0)= \frac{1}{\sqrt{k}} e^{\theta_k} \, , \qquad
\varphi_k'(0)= -i\sqrt{k} \, e^{-\theta_k} \ , 
\ee
where $\theta_k$ is an arbitrary real function\footnote{A potential sign ambiguity in (\ref{cptscalar0}) has been removed by matching the sign with the standard initial conditions for a massless (conformal) field $\varphi_k(0)= \frac{1}{\sqrt{k}}$, 
$\varphi_k'(0)=-i\sqrt{k}$ . Note that we have used the freedom to phase-rotate the modes to make them real and non-negative at $\tau=0$.}.  
In summary, the CPT requirement  reduces the space of possible vacuum states to a  family of states characterized by the hyperbolic initial ($\tau=0$) phase $\theta_k$. In terms of this parameter, the functions $C_{k,1}$ and $C_{k,2}$ read %\footnote{The final expression for $C_{k,1}$ can be written in several ways by using some properties of the gamma functions. In particular we have used $\Gamma(z)\Gamma(z+\frac{1}{2})=2^{1-2z}\sqrt{\pi}\,\Gamma(2z)$ and $\Gamma(z)\Gamma(1-z)=\frac{\pi}{\sin(\pi z)}$.}
\be \label{eq:ck1}
C_{k,1}=2^{i \kappa }\sqrt{\pi}e^{\pi \kappa}\Bigg(\frac{e^{-\frac{i \pi}{4}}e^{\theta_k}}{\kappa^{\frac{1}{4}}\Gamma(\frac{1}{4}-i\kappa)}+\frac{i\,\kappa^{\frac{1}{4}} e^{-\theta_k}}{\Gamma(\frac{3}{4}-i\kappa)}\Bigg)\, ,
\ee
and $C_{k,2}=C_{k,1}^*$.\\

The choice of the vacuum $|0\rangle$ has been then reduced to select a preferred function $\theta_k$. This can be interpreted as giving initial data at $\tau=0$.  Before entering into a detailed discussion of how to single out a preferred vacuum it is convenient to analyze the particle production experienced by late-times observers.\\

\subsection{Particle production and the choice of $\theta_k$}

For each choice of an initial vacuum state $|0\rangle$ we have a different prediction for the particle creation spectrum at late times. Particles are defined at $\tau \to +\infty$  as excitations  of the {\it out-}vacuum $|0_+\rangle$. The particle production rate can be obtained  by the frequency-mixing approach \cite{Parker66}. % It has been  reviewed in \cite{parker-toms, birrell-davies, fabbri-navarro, ford21} and used extensively in the literature for decades (see, for instance, \cite{PF,Mamaev,Audretsch78,Ford, Chung, Anderson-Mottola-Sanders, Good, Ema, BP-F-N-P, Xue}). 
 In our case we find  the following expression  for the average density number of created particles in the mode $\vec k$
\be \label{particle-num-theta}
    n_k \equiv|\beta_k|^2=-\frac{1}{2}+\frac{e^{-\pi \kappa}\cosh{(2 \pi \kappa)}}{4 \pi }\Bigg( e^{-2\theta_k}\,\kappa^{\frac{1}{2}}\,|\Gamma(\tfrac{1}{4}+i\kappa)|^2+e^{2\theta_k} \,\kappa^{-\frac{1}{2}}\,|\Gamma(\tfrac{3}{4}+i\kappa)|^2\Bigg), 
    \ee
    where $\beta_k$ is one of the Bogoliubov coefficients. 
% It is interesting to remark that the above general expression can be reexpressed, after some manipulations, as    \be   |\beta_k|^2=\frac{1}{2}\Big(\cosh(2 \eta_k) \cosh( \Lambda_k)-1\Big)   \ , \ee   where we have defined $\cosh(\Lambda_k)=\sqrt{1+e^{-4 \pi \kappa}}$, and where    \be   2 \eta_k =-2\theta_k +\frac{1}{2} \ln\Big(\frac{\kappa\cosh{(2 \pi \kappa)}}{ 2 \pi^2} |\Gamma(\tfrac{1}{4}+i\kappa)|^4\Big)   \ . \ee   
%Following \cite{Large} 
The number of created particles is minimized when (for more details see \cite{Nadal23}) %$\eta_k=0$. This condition  can be used to choose a preferred CPT-invariant vacuum  and it is characterized by    
\be \label{thetakT}
    \theta_k^{late}=\frac{1}{4}\ln\Big(\frac{\kappa\cosh{(2 \pi \kappa)}}{2 \pi^2} |\Gamma(\tfrac{1}{4}+i\kappa)|^4\Big)\,.
    \ee
%[As we will explain shortly, the super-index $ad$ refers to the fact that this vacuum state is an adiabatic vacuum to all orders]. Therefore, 
In this case we obtain 
\be
n_k=\frac{1}{2}\left(\sqrt{1+e^{-4\pi \kappa}}-1\right)\, .
\ee
For large $k$ it has an exponential decay\footnote{This behaviour is exact for every $k$ for the particle production first considered in \cite{Audretsch78} using the $|0_-\rangle$ as the initial vacuum. } $n_k \sim e^{- \frac{\pi k^2}{\gamma}}$.
We can regard (\ref{thetakT}) as an equivalent characterization of the preferred CPT-invariant vacuum proposed in \cite{LetterCPT, Large}, namely, $|0_T\rangle \propto |0_{+}\rangle + |0_{-}\rangle $. We note that this vacuum also minimizes the energy density in the asymptotic regions  $\tau \to \pm\infty$
\be
\rho_k(\tau)\sim \omega_k|\beta_k|^2
\ . \ee
%{\color{blue}SN: Potser llevar esta frase: where $\beta_k$ refers now to the Bogolubov transformation between $|0_T \rangle$  and  $|0_+\rangle$ or between $|0_T \rangle$  and $|0_-\rangle$  }. {\color{red}(Igual se pot llevar el $|\beta_k|^2$ de l'equació 10 i en esta ficar $n_k$, i no cal definir $\beta$)}

An alternative choice for the vacuum is simply the minimal choice $\theta_k=0$. It corresponds to the conformally invariant solution at $\tau \to 0$ as the equation of motion \eqref{eqm} suggests
\be \label{conformalv}\varphi_k \sim \frac{1}{\sqrt{k}}e^{-ik \tau}\ . \ee
Furthermore, this choice
 minimizes the energy density at $\tau=0$ for $\xi=1/6$. In this case, we have
\be
\rho_k(\tau) \sim  \frac{k}{a^4}\cosh(2 \theta_k)\, .
\ee
For this choice and for large-$k$, the particle number behaves as
\be
n_k \sim k^{-8}\, .
\ee
%It is given by $\theta_k=0$. For this choice, the field modes behave as \be \label{conformalv}\varphi_k \sim \frac{1}{\sqrt{k}}e^{-ik \tau}\ . \ee

\subsection{Ultraviolet  regularity and the Hadamard condition}

For a quantum state $|0\rangle$ to be admitted as physically acceptable we should demand it to be ultraviolet regular.
This means that the high-energy behaviour of the state must approach the behaviour of Minkowski space at a rate such that, {\it at least},  basic composite operators, as the stress-energy tensor, can be renormalized. This is a necessary condition  to properly study the semiclassical Einstein equations. Therefore, it is natural to require  the renormalized form of
$\langle 0 | T_{\mu\nu}|0 \rangle_{ren}$ to be well-defined. Since the renormalization subtractions in evaluating  $\langle 0 | T_{\mu\nu}|0 \rangle_{ren}$ are essentially unique (up to finite counterterms), the ultraviolet behaviour of the modes must  be compatible with the subtractions.  

The mode decomposition (\ref{modecomp}) allows us to write the formal vacuum expectation values of the stress-energy tensor as a sum over modes
\be \label{Tformal}\langle T_{\mu\nu}\rangle = \int d^3k \ T_{\mu\nu}(\vec k, \tau) \ . \ee
A very useful renormalization scheme, which works in momentum space and does not depend on an explicit regulator, was introduced in \cite{parker-fulling1,parker-fulling2,parker-fulling3, Anderson-Parker}. It involves
 a subtraction algorithm directly for the integrand  in (\ref{Tformal}). The method is based on the adiabatic WKB-type expansion of the field modes
\be\label{WKB} \varphi_k (\tau) \sim \frac{1}{\sqrt{W_k(\tau)}} e^{-i\int^\tau W_k (\tau') d\tau'} \ , \ee
where 
\be W_k = \omega_k + \omega_k^{(2)}+ \omega_k^{(4)}+ \cdots \ , \ee
with $\omega_k= \sqrt{k^2 + m^2 a^2}$, and $\omega_k^{(n)}$ is the $n$-order term in a systematic adiabatic expansion. The adiabatic order is fixed by the number of $\tau$-derivatives of the scale factor $a(\tau)$. 
The above expansion of the modes is translated to an adiabatic expansion of  $T_{\mu\nu}(\vec k, \tau)$. This operation allows to extract from the integrand  
$T_{\mu\nu}(\vec k, \tau)$ the divergent part of the overall integral (\ref{Tformal}). Therefore, one can write
\be \label{Tren}\langle T_{\mu\nu}\rangle_{ren} = \int d^3k \ [T_{\mu\nu}(\vec k, \tau)- T^{N-ad}_{\mu\nu}(\vec k, \tau)] \ , \ee
where $N$ is the adiabatic order required to cancel the UV divergences of the integral (\ref{Tformal}). $T^{N-ad}_{\mu\nu}(\vec k, \tau)$ is the associate adiabatic expansion of $T_{\mu\nu}(\vec k, \tau)$ at order $N$. We note that $N$ can be regarded as the degree of divergence of (\ref{Tformal}). For the stress-energy tensor $N=4$ (in four spacetime dimensions).
Since $\nabla^\mu T^{N-ad}_{\mu\nu}(\vec k, \tau) =0$, the subtractions constructed this way preserve  the covariant conservation of the renormalized stress-energy tensor. %The above expression for $\langle T_{\mu\nu}\rangle_{ren}$ can be shown \cite{parker-fulling1} to agree with the result of the method introduced by Zeldovich and Starobinsky \cite{ZS72}. 
%A similar discussion can be performed for other composite operators, like the basic two-point function at coincidence $\langle \phi^2 \rangle$. In this case we have\be \langle \phi^2 \rangle_{ren}= \frac{1}{2(2 \pi)^3 a^2}\int d^3k \ [ |\varphi_k (\tau)|^2 - (|\varphi_k(\tau)|^2)^{N-ad}] \ . \eeHere we need $N=2$.
The  adiabatic subtraction method can also be regarded as a version of the  general DeWitt-Schwinger renormalization method when the spatial translations symmetry of the background is assumed \cite{beltran-nadal1}.  More details can be found in \cite{parker-toms, birrell-davies}.%and a brief sketch in the associated reference \cite{nosotros}.  \\

Since the subtractions $T^{N-ad}_{\mu\nu}(\vec k, \tau)$ are essentially fixed, the major point now is to restrict the field modes involved in   $T_{\mu\nu}(\vec k, \tau)$, defining $|0 \rangle$,  such that $\langle 0 | T_{\mu\nu}| 0  \rangle_{ren}$ be convergent. To this end, and in the context of our analysis, it is useful to evaluate the complete adiabatic  expansion of the modes (\ref{WKB}) at $\tau=0$.  We get 
\be \label{aexpansion}\varphi_k (0) \sim \frac{1}{\sqrt{k}}+\frac{\gamma^2}{8\,k^{9/2}}+\frac{41 \gamma^4}{128\, k^{17/2}}+\cdots\, .\ee
From this we infer the asymptotic expansion of  $\theta_k$ 
\be \label{aexpansiontheta}\theta_k \sim \frac{\gamma^2}{8 k^4}+\frac{5 \gamma^4}{16 k^8}+\frac{61\gamma^6}{24 k^{12}}+\cdots \, . \ee
 The set of vacuum states that fits the above large $k$ expansion up to a given order  can be generically referred to as adiabatic vacuum states of finite order.
  The above {\it adiabatic condition}  at all orders [generically (\ref{WKB}), or   \eqref{aexpansiontheta} in this specific case] is equivalent to the {\it Hadamard condition}, as shown in  \cite{Pirk93}. %Therefore, in a FLRW spacetime, saying that a vacuum state  is a Hadamard state is equivalent to stating that it is an adiabatic state (one assumes, implicitly, that is refereed to an adiabatic state of infinite  order). {\color{blue} Comentar esta ultima frase} \\ % It is not difficult to check that (\ref{thetakT})  defines a vacuum of infinite adiabatic order.  This is so because the large $k$ asymptotic expansion of (\ref{thetakT})  exactly match the adiabatic expansion (\ref{aexpansiontheta}) at any order.\\

A consequence of having a vacuum of {\it infinite adiabatic order}  is that the number density of created particles in the mode $\vec k$ (i.e., $n_k$) with respect to the {\it out}-vacuum (which is also of infinite adiabatic order)  decays  faster than any power $k^{-N}$ for large $k$. Typically, an exponential decay. This very smooth behaviour of the field modes characterizing $|0\rangle$  ensures  not only the ultraviolet convergence of the expectation values of the (renormalized) stress-energy tensor
$\langle 0 | T_{\mu\nu}|0 \rangle_{ren}$, but also fluctuations of it. % and higher-order correlations. 
Furthermore, 
all polynomial functions of the stress-energy tensor, or the existence of Wick polynomials of arbitrary order of the field, require the Hadamard condition \cite{BF2000, hollands-wald-01, hollands-wald-02}. In contrast, for a vacuum of a finite adiabatic order the decay is a power-law. If we choose $\theta_k$ different from  (\ref{aexpansiontheta}) we  get these kind of vacua. For instance, choosing $\theta_k=0$ (zero adiabatic order) we find \cite{nosotros} $n_k \sim k^{-8}$. Furthermore, for $\theta_k \sim 1/k^4$  we also get $n_k \sim k^{-8}$, except for the case $\theta_k = \gamma^2/(8 k^4)$, for which  $n_k \sim k^{-16}$. This behaviour ensures the UV convergence of the total number density of created particles,%\footnote{{\color{blue}The convergence of the total number of created particles ensures that the two representations for the different vacua are unitarily equivalent \cite{Fulling79, Wald94, AgulloAshtekar2015}}},
 \be \label{nk} \int \frac{d^3k}{(2\pi a)^3} n_k \ < +\infty \ . \ee
 In contrast, for $\theta_k \sim 1/k$ we get $n_k\sim k^{-2}$ and the  integral evaluating the total number density is divergent.

It is less clear whether the convergence of (\ref{nk}) is enough to ensure the UV convergence of the  most relevant quantities such as the energy density and pressure $\langle 0 | T_{\mu\nu}|0 \rangle_{ren} $. In Ref. \cite{nosotros} it has been analyzed this issue for the minimal solution  $\theta_k=0$. %The finiteness of the  adiabatic order of the vacuum state originates the emergence of oscillatory terms, or even extra negative powers in $k$, in $T_{\mu\nu}(\vec k, \tau)$. UV regularity of the vacuum state is achieved if those  terms do not nullify the UV convergence of the overall $k$-integral. 
% {\color{blue} In Appendix \ref{ap:large-momentum} } we have analyzed the general case. 
Analyzing the ultraviolet behaviour of the  stress-energy tensor for a general $\theta_k$ (see \ref{ap:large-momentum} for more details), we have found that to ensure its renormalizability,  the asymptotic behaviour of  $\theta_k$ should be  
\be \label{asyms} \theta_k\sim 0 + \mathcal{O}\left(\frac{1}{k^n}\right)  , \ n>3
\, .\ee 

%If we choose $\theta_k$  matching  with the adiabatic expansion (\ref{aexpansiontheta}) up to a given order $k^{-N}$, and assume this expression as valid for arbitrary momenta, we will obtain a vacuum of finite  adiabatic order. The larger is $N$, the faster is the decay rate of the oscillatory terms and the faster is the decay of the particle number at large $k$. In the limit $N \to \infty$ we have vacua  of infinite adiabatic order [one example is (\ref{thetakT})], for which the decay of the particle number is exponential and no oscillatory terms are present in $T_{\mu\nu}(\vec k, \tau)$.

\section{States of low energy}
\label{sec:SLE-scalars}

In general spacetimes, a natural procedure to select and construct a preferred Hadamard vacuum state (i.e. a state of infinite adiabatic order) is not available. 
For FLRW spacetimes the problem is somewhat alleviated, due to the existence of isometries, and the construction of exact Hadamard states is possible \cite{Olbermann07}. These states are defined by minimizing the energy density after averaging with a temporal window function in a smooth region, no matter how short\footnote{ The minimization of the averaged energy density with a temporal window comes from the minimization of the averaged Hamiltonian.  This justify the  factor $\sqrt{|g|}$ in \eqref{eq:smearedED}. For more details see \cite{Nadal23}.}. If the window function shrinks to a point, we recover the traditional  method of instantaneous diagonalization of the Hamiltonian, which produced states that are not Hadamard.  The analysis of \cite{Olbermann07, Banerjee20} was only given for scalar fields (minimally coupled to the curvature). A more general survey, including spin-$\frac{1}{2}$ fields has been given in \cite{Nadal23}. Here we will briefly review the construction for minimally coupled scalar fields and connect with the examples presented above. 

  The starting point is to selected  a fiducial set of normalized modes $\phi_k(\tau)$. They are related to the Weyl modes of the previous section by $\varphi_k= a\, \phi_{k}$. We can choose them by convenience. One can parametrize a general set of modes $T_k(\tau)$ in the form
\be \label{TS}T_k(\tau) = \lambda_k \phi_k(\tau) + \mu_k \phi_k^*(\tau) \ , \ee
where $\lambda_k$ and $\mu_k$ are complex numbers that must obey $|\lambda_k|^2 -|\mu_k|^2 =1$. The goal now is to find for which values of $\lambda_k$ and $\mu_k$ the smeared energy density is minimal. It is given by 
\bea \label{eq:smearedED}
\mathcal{E}_k[f] &:=& \int \text{d} \tau\, \sqrt{|g|}\, f^2 \, \rho_k \, ,
\eea
where $f^2(\tau)$ characterizes the window function and $\rho_k(\tau)$ is given by ($\omega^2=k^2 + m^2 a^2$)
\be \label{modesrhok}
\rho_k(\tau)=\frac{1}{4 a^{2}}\left(|T'_k|^2+\omega^2 |T_k|^2\right)\, .
\ee
Note that $\sqrt{|g|}= a^4$. 
The above formula for $\mathcal{E}_k$ can be reexpressed as
\be
\mathcal{E}_k=(2 \mu_k^2+1)c_{k,1}+2 \mu_k \textrm{Re}(\lambda_k c_{k,2})\, ,
\ee
where we have defined 
\bea \label{eq:c1_conformal_Weyl}
c_{1}&\equiv&c_{k,1}=\frac{1}{4} \int  \text{d} \tau\, \sqrt{|g|}\,\frac{f^2}{a^2}\left(|\phi'_k|^2+\omega^2|\phi_k|^2\right) \, ,\\
c_2&\equiv&c_{k,2}=\frac{1}{4} \int \text{d} \tau \sqrt{|g|}\,  \frac{f^2}{a^2}\left(\phi'_k{}^2+\omega^2 \phi_k^2\right)\, . \label{eq:c2_conformal_Weyl}
\eea
  It can be showed that if we take $\mu_k$ to be real and positive, the minimization problem over the parameters $\lambda_k$ and $\mu_k$ determines a unique solution, namely
\be\label{mu-lambda} \mu_k= \sqrt{\frac{c_{1}}{2\sqrt{c_{1}^2 -|c_2|^2}} -\frac{1}{2}} \ ; \qquad \lambda_k=-e^{-i \text{Arg} \ c_2}  \sqrt{\frac{c_{1}}{2\sqrt{c_{1}^2 -|c_2|^2}} +\frac{1}{2}} \ . \ee
Notice that the minimization problem holds whenever $\frac{|c_2|}{c_{1}}<1$ is satisfied. This  is usually the case if $\phi_k$ do not contain singularities in the support of $f^2$ \cite{Olbermann07, Banerjee20}. This issue will be relevant around the Big Bang singularity, as we will see in the next section. %We remark that the resulting state of low energy is independent of the fiducial solution $\phi_k$ and it is a Hadamard state (i.e., a state of infinite adabatic order).\\

%{\color{red}Simplificar lo que sigue al maximo, siendo coherentes}
%\subsection{CPT-invariant States of Low Energy}
The method can also be extended by constraining the minimization problem to CPT-invariant states, where we are now focusing in a radiation-dominated universe. In this case, we can parametrize the state of low energy using the initial hyperbolic phase $\theta_k$. For this, we choose as a fiducial solution
\be \label{eq:basis-CPT-scalars}
\phi_k(\tau)=\frac{1}{a}\varphi^{CPT}_k(\tau,\theta_k=0)\, , \ee with $\varphi_k^{CPT}$ given in \eqref{sol-scalars} with $C_{k,1}$ and $C_{k,2}$ as in \eqref{eq:ck1} and setting $\theta_k=0$. In this context, $\lambda_k$ and $\mu_k$ can be writen in terms of the initial angle $\theta_k$ as $\lambda_k=\cosh(\theta_k)$ and $\mu_k=\sinh(\theta_k)$, and the smeared energy density results in
\be
 \mathcal{E}_k=c_{1} \cosh(2\theta_k)+\textrm{Re}(c_2)\sinh(2 \theta_k)\, .
\ee
Finally, taking $\partial_{\theta_k} \mathcal{E}=0$ we end up with the low energy CPT-phase
\be\label{CPT-SLE}
\theta_{k}= \frac{1}{2} \text{arctanh}\left(- \frac{\text{Re}(c_2)}{c_{1}}\right)\, .
\ee
Although a particular fiducial solution has been chosen in the minimization process, the final result for $T_k$ is independent of this basis, as proven in \cite{Olbermann07}.  However the resulting state depends, in general, on the choice of $f^2$.

%{\color{red} [Comentar que los ejemplos de vacios dados en la introduccion y despues se pueden ver como estado de low energy en...]} 
The  vacuum states introduced in the previous sections $|0_{\pm}\rangle$ and $|0_{T}\rangle$ can be now re-interpreted as states of low energy for smearing functions with support at  $\tau \sim \pm \infty$  together with the symmetry requirements. Using the general prescription \eqref{mu-lambda} and for any smearing function $f^2$ with support at  $\tau \sim \pm\infty$, we obtain $|0_{\pm}\rangle$. On the other hand, if we also impose CPT  we find that the state of low energy for any window function with support at either $\tau\sim \pm \infty$ is $|0_{T}\rangle \propto |0_{-}\rangle+|0_{+}\rangle$, or, in terms of the initial phase \eqref{CPT-SLE}
\be\theta_k=\theta_k^{late}\, ,\ee
with $\theta_k^{late}$ given in Eq. \eqref{thetakT}. We  recall that, at the asymptotic regions $\tau \to \pm \infty$ the smeared energy density reads (here we include the conventional renormalization subtractions)

\be
\mathcal{E}_k\left[f\right] \propto \frac{1}{2} \int \text{d} \tau\, f^2\,  \omega\,|\beta_k|^2 \,  ,
\ee
so that it is  minimized for the state that produces the minimum amount of particles, i.e., $\theta_k=\theta_k^{late}$. It is important to remark that this state  minimizes the (CPT-symmetric) smeared energy density at late times independently of the choice of smearing function $f^2$. As expected, this state is 
Hadamard. It can be checked by evaluating the asymptotic large $k$ expansion of $\theta_k^{late}$ 
    \be \label{thetakT-late}
    \theta_k^{late}=\frac{1}{4}\ln\Big(\frac{\kappa\cosh{(2 \pi \kappa)}}{2 \pi^2} |\Gamma(\tfrac{1}{4}+i\kappa)|^4\Big) \sim \frac{\gamma^2}{8 k^4}+\frac{5 \gamma^4}{16 k^8}+\frac{61\gamma^6}{24 k^{12}}+\cdots \,  \ee
and comparing with the adiabatic expansion (\ref{aexpansiontheta}). They agree at all orders.  %It agrees with the adiabatic expansion (\ref{aexpansiontheta}) at any order.  
% {\color{blue}SN: Potser llevar esta frase ja que ja està dit dalt de (35) We remark if we do not impose CPT invariance, the states of low energy at early and late times are just $|0_\pm \rangle$, which have zero (renormalized) energy and that cannot be described with the initial phase. }

\section{ States of Low Energy at the Big Bang }\label{Subsec::BigBang::Scalars}

%In the last subsection, we have obtained a CPT-invariant state by minimizing the smeared energy density at  $|\tau| \to \infty$. However, this may seem an unnatural vacuum state since we are imposing an initial condition $\theta_k=\theta_k^{late}$ which is determined by the late time behaviour of the universe. %behaviour of the modes. 
%Therefore, a simple question arises: can we obtain a Hadamard state by minimizing the smeared energy density supported around the Big Bang, $\tau = 0$ ? 

The states described in the last section are natural candidates for the quantum vacuum defined at $\tau \sim \pm \infty$. However, a more physically sensitive choice for the vacuum state would be a state  which minimizes the smeared energy density around the Big Bang. %{\color{blue} cregueu que seria interessant posar en algun lloc un comentari tipus que depenent de si considerem un physical bounce (de $-\tau$ a $\tau$)o si considerem un univers que comença a partir del Big Bang seria físicament més sensat un buit o un altre ? No estic segur d'açò  } 
Therefore, we study here the states of low energy in a CPT-symmetric radiation-dominated universe with $f^2$ supported around the Big Bang ($\tau=0$). In Ref. \cite{Nadal23} it was argued that for $\xi\neq\frac{1}{6}$ an extra condition should be imposed to any window function with support at $\tau=0$, namely 
\be \label{eq:limitf}\lim_{\tau\to 0} \frac{f^2(\tau)}{\tau^2}<\infty\, .\ee
This condition is necessary to ensure that $c_1>|c_2|$ for all $k$, which guarantees a well-behaved state of low energy for all $k$. For simplicity, we use a Gaussian smearing function
\be\label{Gaussian}
f^2=a^2f_g^2\, , \qquad \text{with}\qquad f_g^2= \frac{1}{\sqrt{\pi}\epsilon}e^{-\frac{\tau^2}{\epsilon^2}}\, .\ee

In this section, we will look in detail at the massless and minimally coupled case, relating it to the production of primordial gravitational waves. We will also comment on the generalization for massive scalar particles.% {\color{red}and for spin-$\frac{1}{2}$ massive particles, obtaining an alternative cold dark mater spectrum using a state of low energy around the big bang as the vacuum.}

\subsection{Massless case: primordial gravitational waves}
%{\color{red} Conectar con gravitation Waves. Citar Ford and Parker, 78 \cite{FordParker77}; y Ford review particle creation \cite{ford21}. } {\color{blue} Cal ser cuidadosos en lo dels gravitons, ja que la predicció que ens surt a nosaltres per al numero de gravitons per unitat de volum creat pot ser molt gran ja que va com $1/\epsilon^{3}$. No obstant, diria (discutir amb Pepe) que les observacions d'ones gravitatories primordials (tensor modes) indiquen que són xicotetes. Aleshores no se com pot quadrar aquesta predicció, tot i que no se si es pot relacionar directament el càlcul nostre amb les observacions}\\

The massless and minimally coupled scalar field is of particular interest because it is closely related to the question  of the quantum creation of gravitons. Tensor perturbations can be  quantized in an expanding universe \cite{FordParker77}. One can remove the gauge freedom by fixing the gauge. It is quite common to impose the  transverse trace-free gauge. The two physical degrees of freedom, associated with the  two polarizations, obey a scalar wave equation with $m=0=\xi$. Therefore,  gravitons can be regarded as a pair of massless minimally coupled scalar fields. For massless and conformally coupled  fields there is a natural definition of the vacuum in an expanding universe. This is the conformal vacuum, and the corresponding modes are given by   
\be \label{modesc}\phi_k= \frac{e^{-i k \tau}}{a\sqrt{k}}\, .\ee
The consequence is the absence of particle production. In  a spatially flat radiation-dominated spacetime, the wave equation cannot distinguish between conformal and minimal couplings. This is so because $R=0$. Therefore, one is tempted to also define the vacuum for massless and minimally coupled fields with the modes (\ref{modesc}). Accordingly, one predicts that no gravitons are created by the expansion of the universe during the radiation-dominated phase. One can further argue that the creation of gravitons in the very early universe is a signal of a different  expansion law, such as those linked  to the inflationary universe. We would like to briefly note  that this is a quick conclusion. In the context of our analysis, the natural vacuum for a minimally coupled scalar field is the state of low energy around $\tau\sim 0$. This state coincides with the conformal vacuum for a conformally coupled field, but not for a minimally coupled field.

Let us analyze with more details the massless field with $\xi=0$.
%Let us first analyze the massless case, where we can find analytic solutions. %We will work with the Weyl modes $\varphi_k(\tau)=a(\tau)\phi_k(\tau)$.  
We take the fiducial solution given in \eqref{eq:basis-CPT-scalars}, that in this case corresponds to the conformal solution (\ref{modesc}). 
%\be \phi_k= \frac{e^{-i k \tau}}{a\sqrt{k}}\, .\ee
We proceed to obtain the CPT-invariant state of low energy with the Gaussian function centered at $\tau=0$, satisfying the restriction \eqref{eq:limitf}, namely $f^2=a^2 f_g^2$ [see Eq. \eqref{Gaussian}]. Following the states of low energy prescription, and after some algebra, we obtain the following result
\be 
T_k=\cosh(\theta_k) \,\frac{e^{-i k \tau}}{a\sqrt{k}}+\sinh(\theta_k) \, \frac{e^{+i k \tau}}{a\sqrt{k}}\, ,
\ee
with
\be \label{solutiongravitons}\theta_{k}= -\frac{1}{2} \text{arctanh} \left(e^{-(\epsilon k) ^2}\frac{1+2(\epsilon k) ^2}{1+ (\epsilon k)
   ^2}\right) \, . \ee
 This state is Hadamard because the large momentum expansion of $\theta_k$ decays faster than any power of $k^{-n}$. It is important to stress that this result differs from the conformal vacuum choice $\theta_k=0$. The  implications of the solution (\ref{solutiongravitons}) merits  further analysis and it is out of the scope of this work.  
%The divergent terms $\tau^{-2}$ in the integrands of Eqs. \eqref{eq:c1_c2_conformal_Weyl_Massless1} and \eqref{eq:c1_c2_conformal_Weyl_Massless2} disappear for a conformally coupled field (see Appendix \ref{App::SLE::Conf}) and the $\sqrt{|g|}=a^{4}$ factor from the volume element is enough to render both integrals finite. However, because we are minimizing the state around the Big Bang, the resulting state depends on the test function $f^2$.

\subsection{Massive case %and spin-$\frac{1}{2}$ fields
}
We briefly describe here the minimization problem for massive scalar fields.
%For completeness, we can also comment on the minimization problem for massive particles. 
A more detailed analysis is given in Ref. \cite{Nadal23}. As before, one can choose the upgraded Gaussian smearing function $f^2=a^2f_g^2$ centered at $\tau=0$, and the fiducial CPT-invariant solution $\phi_k$ given in \eqref{eq:basis-CPT-scalars} to obtain the state of low energy centered at $\tau=0$. In this context,  the $\tau$-integrals for $c_1$ and $c_2$  given in Eqs. \eqref{eq:c1_conformal_Weyl} and  \eqref{eq:c2_conformal_Weyl} cannot be analytically resolved. However, making an expansion of the mode functions $\phi_k$ around $\tau=0$ it is possible to obtain an approximate result which is enough to check, order by order, the large $k$ behaviour of the resulting state of low energy. Computing a large $k$ expansion of the associated low-energy phase $\theta_k$ defined in Eq. \eqref{CPT-SLE}, it was checked that it obeys the expected adiabatic expansion \eqref{aexpansiontheta} up to a given order, that increases as the order of the expansion in powers of $\tau$ is increased.  In particular, it was checked that for $\mathcal{O}(\tau^{14})$ the  large momentum expansion matches with the adiabatic expansion up to $ \mathcal{O}( k^{-16} )$, namely\be \label{thetakbigbang}
    \theta_k \sim \frac{\gamma^2}{8 k^4}+\frac{5 \gamma^4}{16 k^8}+\frac{61\gamma^6}{24 k^{12}}+ \mathcal{O}( k^{-16}) \, . \ee

\section{Fermions}\label{sec:fermions}
%{\color{red}describir de palabra los resultados para fermiones, dar la formula de creación de partículas, las predicciones para la masa de los neutrinos, la condición asintónica para $\Theta_k$, la condición para renormalizar $T_{\mu \nu}$ etc...}
The analysis in Sections \ref{sec:scalars-CPT} and \ref{sec:SLE-scalars} can be extended for spin-$\frac{1}{2}$ fields. In this case, we can study the time evolution of the field in terms of two time-dependent mode functions $\varphi_k \to \{h^{I}_k,h^{II}_k\}$ (see Ref. \cite{Nadal23} for the details) that obey
\bea \label{eq:modeh1}
h_k^{\prime I}+i k h_k^{I I}+i \gamma \tau \, h_k^I&=&0, \\
h_k^{\prime I I}+i k h_k^I-i \gamma \tau\,  h_k^{I I}&=&0\, , \label{eq:modeh2}
\eea
together with the normalization condition $|h_k^I|^2+|h_k^{I I}|^2=1$.
 The condition for a CPT-invariant vacuum state takes the following simple form 
\be
h^{I}_k(\tau)=h^{II*}_k(-\tau)\, ,
\ee
and hence, any CPT-invariant vacuum state can be parametrized by an initial phase $\Theta_k$ as
\be
h_k^I(0)  =\frac{e^{+i \Theta_k}}{\sqrt{2}} \, , \qquad
h_k^{I I}(0)  =\frac{e^{-i \Theta_k}}{\sqrt{2}}\, .
\ee
As in the scalar case, we can also compute the particle creation at late times for the vacuum state $|0\rangle$, which is characterized by the trigonometric initial phase $\Theta_k$. The vacuum $|0\rangle$ is perceived at late times as a collection of particles, defined as quantum excitations of the adiabatic out-vacuum $\left|0_{+}\right\rangle$. It is found  
\be \label{eq:nk-fermions}
n_{k,h}= |\beta_{k}|^2=\frac{1}{2}-\frac{e^{-\pi \kappa}\sinh(2\pi \kappa) \sqrt{\kappa}}{4\pi}\left(e^{- 2 i \Theta_k}e^{i\frac{\pi}{4}}\Gamma(i \kappa)\Gamma(\tfrac{1}{2}-i \kappa)+ e^{2 i  \Theta_k}e^{-i\frac{\pi}{4}}\Gamma(-i \kappa)\Gamma(\tfrac{1}{2}+i \kappa)\right)\ , 
\ee
 where the subindex $h$ refers to the helicity of the created fermionic particles. $\beta_k$ is again one of the Bogoliubov coefficients, and turns out to be independent of $h$. %Note that \eqref{eq:nk-fermions} is independent of $h$. 
 It can be shown that the number of created particles is minimized for 
\be
\Theta_k=\Theta_k^{late}\equiv  \frac{\pi}{8}+\frac{1}{2} \textrm{Arg}[\Gamma(\tfrac{1}{2}-i\kappa)\Gamma(i\kappa)] \, 
\ ,
\ee
resulting in 
\be
n_{k,h}=\frac{1}{2}\left(1-\sqrt{1-e^{-4 \pi \kappa}}\right)\, .
\ee
This quantity decays exponentially for large $k$. The state $\Theta_k^{late}$ agrees with the preferred CPT-invariant fermionic vacuum proposed in \cite{LetterCPT,Large}, that is $|0_T\rangle \propto |0_{-}\rangle+|0_{+}\rangle$. As for the scalar case, an alternative tempting vacuum could be the minimal choice \be \Theta_k=0\, .\ee For this choice,  the field modes behave as the conformal modes for $\tau \sim 0$, namely
\be \label{eq:conformal-dirac}
h^{I,II}_k\sim \frac{1}{\sqrt{2}}e^{-ik\tau}\, ,
\ee
and the particle number exhibits a power-law decay  $n_{k,h}\sim k^{-4}$.

A careful analysis shows that this minimal choice is not compatible with renormalization: for $\Theta_k=0$, the vacuum expectation value of the stress energy tensor $\langle T_{\mu \nu}\rangle_{ren} $ is divergent after renormalization. This can be proved by using the cumbersome techniques displayed in \cite{nosotros}. %Using the adiabatic expansion of the fermionic field modes \cite{LNT1,LNT2} it is found that the required asymptotic (large $k$) behaviour of the initial phase $\Theta_k$ to have a finite vacuum expectation value of the stress-energy tensor after renormalization is 
 Using the adiabatic expansion of the fermionic field modes \cite{LNT1,LNT2} it is found that in order to have a finite vacuum expectation value of the stress-energy tensor after renormalization the required asymptotic (large $k$) behaviour of the initial phase $\Theta_k$ is 
\be\label{eq:condSET:fermions}
\Theta_k \sim-\frac{\gamma}{4 k^2}+\mathcal{O}({k^{-n}}), \qquad n>2\, .
\ee
  We give more details of this result in \ref{ap:large-momentum-spin12}. Moreover, for a vacuum state obeying the adiabatic condition (or equivalently, the Hadamard condition), we require the following asymptotic expansion for the initial phase,
\be \label{eq:asympt:fermions}
\Theta_k \sim-\left(\frac{\gamma}{4 k^2}+\frac{\gamma^3}{6 k^6}+\frac{4 \gamma^5}{5 k^{10}}+\cdots\right) \, .
\ee
%The above adiabatic condition (at all orders) is equivalent to the Hadamard condition.

\subsection{States of low energy}
\begin{figure}[h]
    \centering
 \includegraphics[width=115mm]{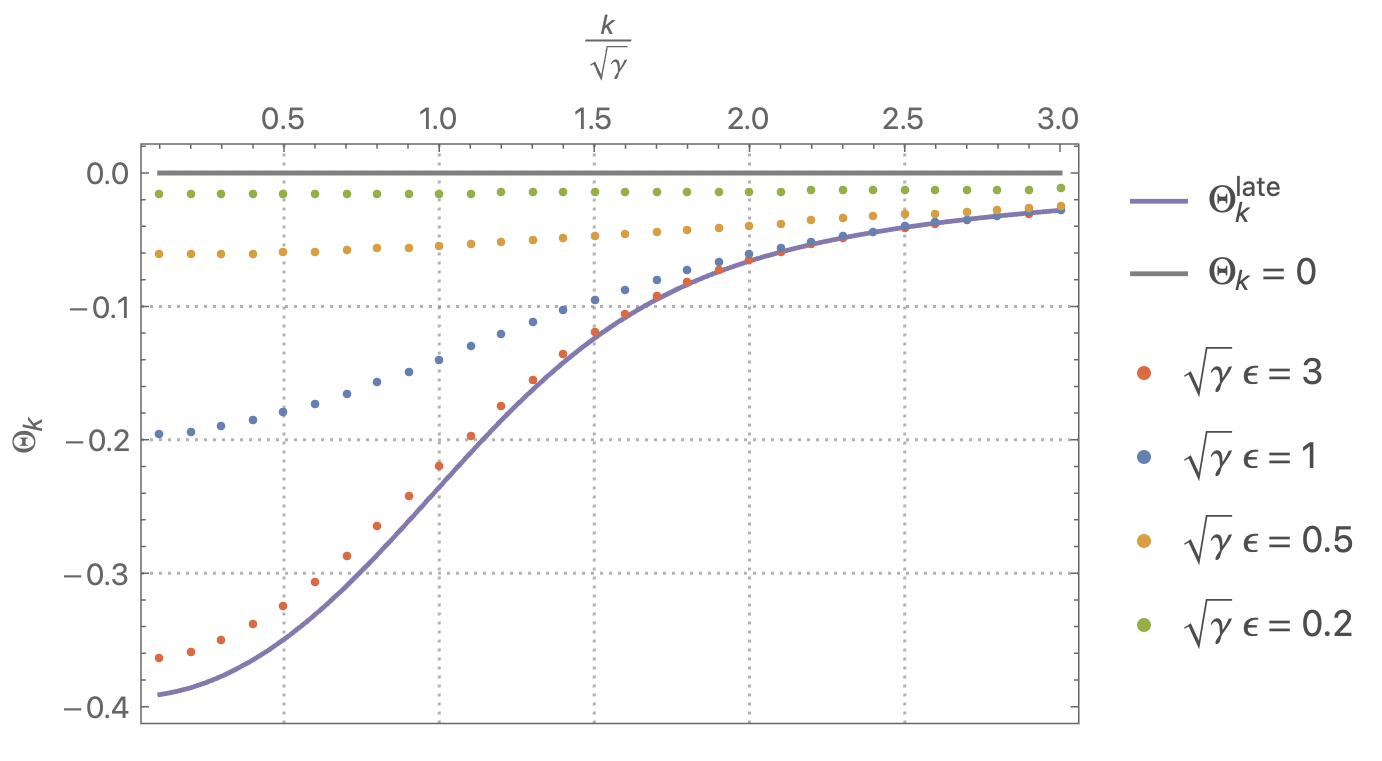} %\hspace{1cm} %\includegraphics[width=70mm]{logplot.png} 
    \caption{ {\small We represent $\Theta_k$ obtained with the states of low energy prescription for different  values of $\epsilon$,  $\sqrt{\gamma}\epsilon=3$ (red dots) and $\sqrt{\gamma}\epsilon=1$ (blue dots), $\sqrt{\gamma}\epsilon=0.5$ (orange dots) and $\sqrt{\gamma}\epsilon=0.2$ (green dots). We also represent the two limiting cases $\Theta_k=0$ (grey line) and $\Theta_k=\Theta_k^{late}$ (purple line).}}
    \label{fig:theta}
\end{figure}

As in the scalar case, it is possible to construct Hadamard states by minimizing the smeared energy density over a temporal window function $f^2$, 
\be
\label{eq:smearedED2}
\mathcal{E}_k[f] := \int \text{d} \tau\, \sqrt{|g|}\, f^2 \, \rho_k \, ,
\ee
where now $\rho_k$ corresponds to the energy density associated with a pair of modes $\{t^{I}_k,t^{II}_k\}$. A detailed analysis of this procedure is given in Ref. \cite{Nadal23}. In particular, it has been determined that upon imposition of CPT-invariance, the state that minimizes the smeared energy is the state corresponding to the initial phase 
\be
\Theta_k=\frac{1}{2} \arctan \left(\frac{\operatorname{Im}\left(c_2\right)}{c_1}\right)\, ,
\ee
where 
\bea
 c_1&=&2 i \int \mathrm{d} \tau f^2\left(h_k^I \frac{\partial h_k^{I *}}{\partial \tau}+h_k^{I I} \frac{\partial h_k^{I *}}{\partial \tau}\right)\,, \\
 c_2&=&2 i \int \mathrm{d} \tau f^2\left(h_k^I \frac{\partial h_k^{I I}}{\partial \tau}-h_k^{I I} \frac{\partial h_k^I}{\partial \tau}\right) \,,
\eea
with $\{h^{I}_k,h^{I}_k\}$  being the (CPT-invariant) solutions to the mode equations \eqref{eq:modeh1} and \eqref{eq:modeh2} with initial conditions $h^{I}_k(0)=h^{II}_k(0)=1/\sqrt{2}$. In parallel to the scalar case, it is easy to check that for any window function $f^2$ with support at $\tau \to \pm \infty$, the CPT-invariant state that minimizes the late times smeared energy density is 
\be
\Theta_k=\Theta^{late}_k\, ,
\ee
 since it minimizes the late-times particle number. Furthermore, it is also possible to minimize the smeared energy density around the big bang $\tau=0$. Without loss of generality, it can be done with a Gaussian function\footnote{We note that for fermions
 we do not need to impose any extra condition to the smearing function. For the scalar case we required the  condition \eqref{eq:limitf}.}
 \be
 f^2=\frac{1}{ \sqrt{\pi}\epsilon}e^{-\frac{\tau^2}{\epsilon^2}}\, .
 \ee
 For massless particles, we obtain $\Theta_k=\Theta_k^{late}=0$, which corresponds to the conformal vacuum \eqref{eq:conformal-dirac}. We note that in this case $n_k=0$ and particles are not created by the expansion of the universe. For massive particles the result is qualitatively different. In this case it is not possible to find an exact analytical result. However,  in \cite{Nadal23}  it has been checked that the asymptotic behaviour of the resulting state is compatible with the Hadamard condition, i.e., for large $k$ it follows \eqref{eq:asympt:fermions}.  Moreover, as for the scalar, case we can obtain an approximated analytical value of $\Theta_k$ by expanding the mode part of the integrals $c_1$ and $c_2$ around $\tau =0$ up to the desired order. %The resulting state for $\Theta_k^{SLE}$ can be used to study particle creation effects}. 
 In order to study particle creation in the next section, it is interesting % In order to compare physical predictions, it is more interesting though 
 to see how the initial phase $\Theta_k$ behaves in the infrared. To this end, we compute $\Theta_k$ numerically for different values of $\epsilon$. The results are represented in Figure \ref{fig:theta}.  We observe that the narrower the Gaussian, the closer the state is to $\Theta_k=0$.  In the limit $\epsilon \to 0$, 
 \be
 f^2=\delta(\tau)\, ,
 \ee
 and the resulting state of low energy is $\Theta_k=0$, which is not Hadamard nor gives a renormalized vacuum expectation value of the stress-energy tensor, as discussed around Eq. \eqref{eq:condSET:fermions}. Conversely, the wider the Gaussian width, the closer is the state to $\Theta_k=\Theta_k^{late}$. This is so because the dominant contribution of the Gaussian window comes from the asymptotic regions $\tau \sim \pm \infty$.

In the next section we will illustrate how the above construction can be applied to a cosmological particle creation process of special relevance. It involves the production of heavy right-handed neutrinos by the  radiation-dominated universe. To motivate the selection of this spin-$\frac{1}{2}$ field as a natural candidate for dark matter we will argue on the basis of conformal symmetry and anomaly cancellation.   

\section{Quantum stress-energy tensor and conformal symmetry}

%This vacuum state turns out to be UV regular. For fermions

The (renormalized) vacuum expectation values of the stress-energy tensor in a FLRW spacetime can be expressed in  the form \cite{parker-toms, DNT}
\be  \langle T_{\mu\nu} \rangle = \langle T_{\mu\nu} \rangle_0 + t_{\mu\nu} \ . \ee
The tensor $\langle T_{\mu\nu} \rangle_0$ is traceless and it vanishes for conformal fields in the conformal vacuum.   $t_{\mu\nu}$ is a geometric tensor accounting for the  trace/conformal anomaly.  It can be univocally derived from the trace anomaly and  expressed as a linear combination of the geometric tensors $^{(3)}H_{\mu\nu}= \frac{1}{12}R^2g_{\mu\nu} -R^{\rho\sigma}R_{\rho\mu\sigma\nu}$ and $^{(2)}H_{\mu\nu}$. [In FLRW spacetimes $^{(1)}H_{\mu\nu}$ is proportional to $^{(2)}H_{\mu\nu}$ and $^{(3)}H_{\mu\nu}$ is a conserved tensor \cite{birrell-davies}].

 As we approach back to the Big Bang, the stress-energy tensor is fully dominated by the conformal anomaly contribution $t_{\mu\nu}$. However, while the metric tensor 
\be ds^2 = a^2(\tau) (d\tau^2 - d\vec x^2) \ , \ee
is conformally regular 
as $\tau \to 0$, the conformally reescaled stress-energy tensor, in particular $a^2(\tau) t_{\mu\nu}$, is not longer regular at $\tau=0$. This is in tension with the basic idea behind  the Weyl curvature hypothesis. %In other words, due to the trace anomaly the initial singularity can not be fully absorbed for $\langle T_{\mu\nu} \rangle$ by a local conformal transformation.

To solve the conflict one should require an exact cancellation of the conformal anomaly in  the Standard Model. In a first approximation, we can ignore interactions and consider the free field content of the Standard Model living in a curved spacetime.  
%\bea g_{\mu\nu} &\to & \bar g_{\mu\nu}=\Omega^2(x) g_{\mu\nu} \nonumber \\
%\langle T_{\mu\nu} \rangle &\to &  \langle \bar T_{\mu\nu} \rangle= \Omega^2 (x) \langle T_{\mu\nu} \rangle \ . \eea
%The argument now is as follows. 
The general expression for the conformal anomaly is given by 
\be  \langle  T^{\mu}_{\mu}  \rangle  \equiv t^\mu_\mu= c C^2 -a E \ , \ee
where $C^2$ is the square of the Weyl tensor and $E$ is the Euler density. We have omitted the contribution proportional to $\Box R$ since it is intrinsically ambiguous \cite{Wald94}. We are allowing geometries beyond exact FLRW, so $C^2$ does not need to be exactly zero.  The relevant point here is that the coefficients $a$ and $c$, which depend on the spin of the field \cite{birrell-davies}, 
\bea a &=& \frac{1}{360(4\pi)^2} [ n_0 + \frac{11}{2} n_{1/2} + 62 n_1 ] \ >0 \nonumber \\
 c &=& \frac{1}{120(4\pi)^2} [ n_0 + 3n_{1/2} + 12 n_1] \ > 0 \ , \eea
cannot  cancel out. All  conventional fields in the Standard Model contribute with the same sign. [$n_0 \equiv$ number of scalar fields;
$n_{1/2} \equiv$ number of spin-$1/2$ Weyl fields;
$n_1 \equiv$ number of spin-$1$ fields]
For a recent discussion on the contribution of Weyl fermions  see \cite{Abdallah:2021eii}.

The cancellation of the conformal anomaly requires the introduction of a special scalar field $\phi'$. Its action is expressed in terms of the unique conformally-invariant fourth order operator 
 \be  S= \frac{1}{2} \int d^4x \sqrt{-g} \phi' \triangle_4 \ \phi ' \ , \ee 
where $\triangle_4$ is given by \cite{Paneitz, Reigert}  (for other spin fields, see \cite{Shapiro, Zanusso})
 \be  \triangle_4 = \Box^2 + 2 R^{\mu\nu}\nabla_\mu \nabla_\nu - \frac{2}{3}R \Box + \frac{1}{3}(\nabla^\mu R) \nabla_\mu \ . \ee
The contribution of this dimensionless scalar field $\phi'$ to the conformal anomaly is given by \cite{Gusynin:1989ky} 
\be a= -\frac{28}{360(4\pi)^2}  \ \ \ \ \ \  c= -\frac{8}{120(4\pi)^2} \ . \ee
The conformal anomaly is exactly cancelled for \cite{Boyle:2021jaz, Miller:2022qil}
$$ n_{1/2} = 4n_1= 48 \ , \ \ \ \ \ \ \ \ \  n'_0 = 3n_1= 36 \ , \ \ \ \ \ \ n_0 =0\, . $$
This corresponds to the current  Standard Model, with exactly three generations, but with the following important caveats: i) one must include right-handed neutrinos, ii) the Higgs should be excluded as a fundamental field (it should emerge as  a composite field), iii) one should also include  $36$ scalar fields $\phi'$. These special fields are very peculiar because they do not have particle excitations. They only contribute to the vacuum state \cite{BogolubovBook}. %In this context, there is a natural candidate for dark matter. A heavy right-handed (sterile) neutrino $\nu^R$, decoupled from the other fields of the Standard Model. 

\subsection{Dark matter candidate  for a time reversal and conformally invariant Big Bang} 

%{\color{blue} Potser posaria lo de la massa del neutrí  en una secció final o en  discussion  perquè la massa del neutrí per al cas $\Theta_k=0$ és la mateixa que obtindriem de l'estat de low energy centrat al Big Bang per a una window function amb amplada molt xicoteta $\Theta^{SLE}_k|_{\epsilon \to 0}\sim -\frac{\gamma \epsilon^2}{4} +O(\epsilon^3)$. Aleshores, tot i que en aquest cas no es renormalitzable, dona una bona predicció per a la massa del neutrí. Si parlem d'açò després d'introduir els estats de low energy podem justificar la massa per al cas natural $\Theta_k=0$. Açò encaixa en el fet que per a fermions $\Theta_k=0$ minimitza la densitat d'energia en $\tau=0$ }.\\

In the above  context  there is a natural candidate for dark matter. A heavy right-handed majorana neutrino $\nu^{R}$, decoupled from all of the other particles in the Standard Model. It can be only produced gravitationally \cite{LetterCPT}. Different initial phases produce then different cold dark matter spectra \eqref{eq:nk-fermions}. 

\begin{figure}[h]
    \centering
\includegraphics[width=120mm]{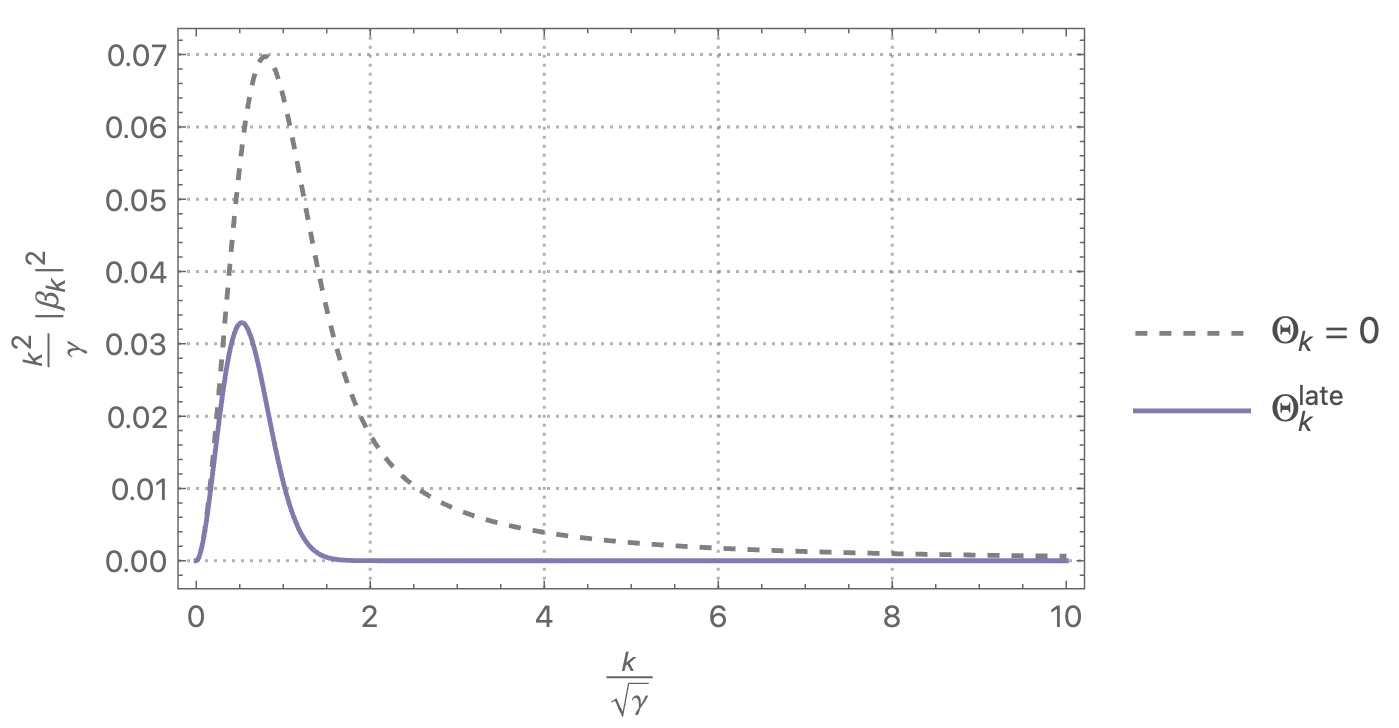}
    \caption{{\small Comparison of the predicted spectra  (cold dark matter) for two different vacua, namely  $\Theta_k^{late}$ and $\Theta_k=0$.}}
    \label{fig:DM}
\end{figure}

It is interesting to remember that for the Hadamard state of low energy centered at the Big Bang, we can consider a very small window function $\epsilon \to 0$. As it was shown in Fig. \ref{fig:theta}, for $\epsilon \to 0$ the SLE tends to $\Theta_k =0$ in the infrared regime. This can also be seen for the approximated SLE if we expand the analytical expression for a very small window function, $\Theta^{SLE}_k|_{\epsilon \to 0}\sim 0+O(\epsilon^2)$. We want to remark this fact because the number density of created particle receives its major contribution from the infrared regime. Therefore, almost the same number density of created particles is obtained by using $\Theta_k=0$ or by using the state of low energy centered at the Big Bang with a very small window function. In Figure \ref{fig:DM} we compare the prediction for $\Theta_k=\Theta_k^{late}$ and $\Theta_k=0$. %and $\Theta_k=\Theta_k^{SLE}$. Here, $\Theta_k^{SLE}$ is computed by expanding the mode  part of $c_1$ and $c_2$ around the Big Bang to $O(\tau^{14})$.
In the following, we will compute the number density of created particles as a function of the state $\Theta_k$, and with this, we will obtain the resulting mass of the dark matter right handed neutrino for the states $\Theta_k= \Theta_k^{late}$ and $\Theta_k =0$.

%\begin{figure}[h]
   % \centering
 %\includegraphics[width=100mm]{plot2.png}
    %\caption{ Comparison of the predicted spectra  (cold dark matter) for different vacua.   $\Theta_k^{SLE}$ is computed with $\epsilon\to0.2$}
   % \label{fig:DM}
%\end{figure}

%In this context, it is interesting to compare the predictions for particle production for two different examples of CPT-invariant vacua.
From the results of Section \ref{sec:fermions}, the comoving number density of created right-handed neutrinos 
is given by
\be n_{dm}= \sum_h \int \frac{d^3k}{(2\pi)^3} |\beta_k|^2 = (\gamma/\pi)^{3/2}I_{\Theta_k}\, .\ee
 $I_{\Theta_k}$ is the dimensionless integral given by
\be I_{\Theta_k}= \frac{4}{\sqrt{\pi}} \int_0^\infty d \kappa \sqrt{\kappa } |\beta_{k}|^2 \, , \ee
where $|\beta_{\kappa}|^2$ is given by the right hand side of \eqref{eq:nk-fermions}. 
Furthermore, the physical energy density of the cold dark matter created is diluted in cosmological time $t$ according to 
%\be  \rho_{dm}(t) = \frac{m_{dm}}{a^3}\sum_h \int \frac{d^3k}{(2\pi)^3} |\beta_k|^2 = \frac{m_{dm}}{a^3}(\gamma/\pi)^{3/2}I_{\Theta_k} = \frac{m_{dm}}{(2\pi)^{3/2}} \left(\frac{m_{dm}}{t}\right )^{3/2} I_{\Theta_k} \, .\ee

\be  \rho_{dm}(t) = \frac{M_{\Theta_k}}{a^3}\sum_h \int \frac{d^3k}{(2\pi)^3} |\beta_k|^2 = \frac{M_{\Theta_k}}{a^3}(\gamma/\pi)^{3/2}I_{\Theta_k} = \frac{M_{\Theta_k}}{(2\pi)^{3/2}} \left(\frac{M_{\Theta_k}}{t}\right )^{3/2} I_{\Theta_k} \, .\ee
To obtain a prediction for $M_{dm}$ from the observed value of the energy density of dark matter we follows now the argument in \cite{Large}. For $\Theta_k^{late}$ used in \cite{Large} one gets $I_{\Theta_k^{late}}=0.01276$ and then $M_{\Theta_k^{late}}= 4.8 \times 10^{8} GeV$. If we now change the quantum state, the prediction for the mass changes accordingly
%\be
%m_{dm}^{5/2}I_{\Theta_k^{late}}=M_{dm}^{5/2}I_{\Theta_k}
%\ee

\be
M_{\Theta_k^{late}}^{5/2}I_{\Theta_k^{late}}=M_{\Theta_k}^{5/2}I_{\Theta_k} \, .
\ee
For $\Theta_k=0$ we have $I_{\Theta_k=0}=0.06305$ and hence  $M_{\Theta_k=0}= 2.6 \times 10^{8} GeV$.

%\end{itemize}

\section{Conclusions}
A straightforward realization of the Weyl curvature hypothesis has been considered by assuming an early universe dominated by radiation and extending the expansion factor analytically to negative values of conformal time. In this context, the problem of selecting a preferred vacuum state is discussed by enforcing the emergent symmetry of the metric under the transformation $\tau \to -\tau$. % as suggested in \cite{LetterCPT}. 
 One possible vacuum choice, as suggested in \cite{LetterCPT}, is the $T$-invariant linear combination $|0_T\rangle\propto |0_+\rangle+|0_-\rangle$, where $|0_+\rangle$ and $|0_-\rangle$ are the adiabatic vacua at $\tau \to \pm \infty$, respectively. %However, this is somewhat a teleological choice since it requires the imposition of initial data {\color{blue} comentar frase} from the entire history of the evolving universe.  

An alternative option is to define a vacuum state around $\tau\sim 0$. This option is more involved and can be formulated using the states of low energy prescription. In this framework, the vacuum state is defined as the state that minimizes the energy density averaged over a temporal window function $f^2$. It is important to emphasize that this approach yields vacuum states that satisfy the Hadamard condition, which is crucial to obtain finite quantities for vacuum expectation values of the stress-energy tensor, and its fluctuations, after renormalization.

The states obtained with this method allow some interesting physical predictions. First, primordial gravitational waves can be produced even though the initial state of the universe is well-described by a radiation-dominated era. Second, the mass of the natural candidate for dark matter in this scenario, namely, a heavy right-handed Majorana neutrino $\nu^R$ is of order $10^{8}GeV$. The specific prediction for the mass slightly depends on the selection of the vacuum state. %  can be up to almost two times smaller than the mass predicted from $|0_T\rangle$. %first predicted in Ref. \cite{LetterCPT}...

\section*{Acknowledgements}
We thank  the organizers of the  conference  ``Avenues of Quantum Field Theory in Curved Spacetime'' [Genoa, Italy; September 14-16, 2022]  for a very inspiring  meeting. An earlier version of this work was presented there.  This work is supported by the Spanish Grants %FIS2017-84440-C2-1-P funded by MCIN/AEI/10.13039/501100011033 “ERDF A way of making Europe”, Grant
 PID2020-116567GB-C2-1  funded by MCIN/AEI/10.13039/501100011033, 
 and  PROMETEO/2020/079 (Generalitat Valenciana). 
P. B. is supported by the Ministerio de Ciencia, Innovaci\'on y Universidades, Ph.D. fellowship, Grant No. FPU17/03712.
 S. N. is supported by the Universidad de Valencia, within the Atracci\'o de Talent Ph.D fellowship No. UV-INV- 506 PREDOC19F1-1005367. S. P. is supported by the Leverhulme Trust, Grant No. RPG-2021-299.

\appendix

\section{Details on the large momentum expansion of the scalar field modes } \label{ap:large-momentum}

This short appendix gives supplementary material to the analysis given in \cite{nosotros} for more general hyperbolic phases $\theta_k$. 
For an arbitrary value of $\theta_k$ we cannot ensure that the stress-energy tensor is renormalizable with the standard subtractions. We have to analyze the problem with the technique displayed in \cite{nosotros}. In short, we have to look at the large $k$ expansion of the modes $\varphi_k$ %(for this specific analysis it is preferable to work in the comoving time $t$)\\
%{\color{red} canviar expressió a:
%$$ |\varphi_k(\tau)|^2= \frac{1}{a(\tau)}|h_k(t)|^2\left|_{t=\frac{a_0^2}{4}\tau^2$$
%Dubte: Posar expressió o posar plot numèric amb diferents $\theta_k$ ?}

\bea
|\varphi_k (\tau)|^2 &\sim & \cosh (2 \theta_k ) \left(\frac{1}{k}-\frac{\gamma ^2 \tau ^2}{2
   k^3}+\frac{3 \gamma ^4 \tau ^4}{8 k^5} +\frac{\gamma ^2}{4 k^5} -\frac{\gamma ^2
   \cos (2 k \tau )}{4 k^5}+  \mathcal{O}\left( \frac{1}{k^{6}}\right)\right) \\ \nonumber
   &&+ \sinh (2 \theta_k) \left(-\frac{\gamma ^2}{4 k^5} +\frac{\cos (2 k \tau )}{k}-\frac{\gamma
   ^2 \tau ^3 \sin (2 k \tau )}{3 k^2} \right. \\
   && \left.-\frac{\gamma ^4 \tau ^6 \cos
   (2 k \tau )}{18 k^3}-\frac{\gamma ^2 \tau ^2 \cos (2 k \tau )}{2 k^3} +  \mathcal{O}\left( \frac{\sin(2k\tau)}{k^{4}}\right) \right)  \nonumber
   %&& \frac{\gamma ^6 \tau ^9 \sin (2 k \tau )}{162 k^4}+\frac{13 \gamma ^4 \tau ^5 \sin (2 k
   %\tau )}{60 k^4}+\frac{\gamma ^2 \tau  \sin (2 k \tau )}{2 k^4} \nonumber \\
   %&& \left. \frac{\gamma ^8 \tau ^{12} \cos (2 k \tau )}{1944 k^5}+\frac{2 \gamma ^6 \tau ^8 \cos
   %(2 k \tau )}{45 k^5}+\frac{13 \gamma ^4 \tau ^4 \cos (2 k \tau )}{24
   %k^5}+\frac{\gamma ^2 \cos (2 k \tau )}{4 k^5}\right)
\ . \eea

To get a finite $ \langle 0 | T_{\mu\nu}| 0 \rangle_{ren}$ the non-oscillatory terms  must coincide with the adiabatic expansion obtained from \eqref{WKB} up to and including the order $k^{-5}$.
These are exactly the non-oscillatory terms inside the parenthesis multiplying to $\cosh (2\theta_k)$. Therefore, the factor $\cosh (2\theta_k)$ can only add corrections of order $\sim k^{-5}$. Taking into account the Taylor expansion of the hyperbolic cosine, we can easily conclude % Using the expansion $\cosh{x}= 1+\frac{x^2}{2}+\frac{x^4}{24}+O\left(x^5\right)$,  we must require 
that the leading order in the ultraviolet expansion of $\theta_k$ must be at least of order $k^{-5/2}$. On the other hand, the leading  oscillatory term must also decay, at least, as $k^{-n}, n >4$.  %i.e$\sim \cos (2 k \tau )/ k^5$.
This ensures a well  behaved integral for the energy density and pressure in the ultraviolet regime, (see \cite{nosotros}). Therefore the factor $\sinh(2\theta_k)$ can only add corrections of order $k^{-n}, n>3$. Using the expansion of the hyperbolic sine, %for the hyperbolic function $\sinh{x}|_{x\to0}= x+\frac{x^3}{6}+\frac{x^5}{120}+O\left(x^7\right)$, 
we see that the leading order in the ultraviolet expansion of $\theta_k$ must be at least of order $k^{-n}, n>3$. Therefore the asymptotic form of $\theta_k$ should be $\theta_k\sim   \mathcal{O}(k^{-n}) ,\, n>3$. \\

 \section{Details on the large momentum expansion of the spin-$1/2$ field modes } \label{ap:large-momentum-spin12}

In this brief appendix, we give the large $k$ expansion of the integrand of the formal vacuum expectation value of the energy density $\rho_k(\tau)$, defined as
\be \langle T_{00} \rangle = \frac{1}{(2\pi)^2 a^3} \int_0^\infty k^2 dk \,\rho_k ( \tau) \ , \ee
for the CPT-invariant vacuum states characterized by $\Theta_k$ (see Ref. \cite{DNT} for details regarding $\langle T_{\mu \nu} \rangle_{ren}$ for Dirac fields in spatially flat FLRW backgrounds and \cite{Nadal23} for the explicit form of the modes in the CPT-invariant radiation-dominated universe). Using the techniques displayed in \cite{nosotros} we find
\bea \label{eq:dirac-largek}
m^{-1}\rho_k(\tau)&\sim &  \cos(2 \Theta_k)\Bigg(-\frac{2k}{\gamma \tau}-\frac{\gamma \tau}{k}+\frac{\gamma}{2 k^3 \tau}+\frac{\gamma^3 \tau^3}{4 k^3}-\frac{\gamma \cos(2 k \tau)}{2 k^3 \tau}+\mathcal{O}\left(\frac{\sin(2 k \tau)}{k^4}\right)\Bigg)\\
&&+\sin(2 \Theta_k)\Bigg(+\frac{1}{k \tau}+\frac{\gamma^2 \tau}{2k^3}-\frac{\cos(2k \tau)}{k \tau}+\frac{\gamma^2 \tau^2 \sin(2k\tau)}{3k^2}+\mathcal{O}\left(\frac{\cos(2 \kappa\tau)}{k^3}\right)\Bigg)\,.\nonumber
\eea
This expression has to be compared with the large $k$ expansion of the adiabatic expansion of the energy density%up to and including the second adiabatic order (the 4th adiabatic order is finite for spin-$1/2$ fields)
, namely
\be \label{eq:ad-largek}
m^{-1}\rho_k(\tau)^{ad}\sim -\frac{2 k}{\gamma \tau}-\frac{\gamma \tau}{k}+ \frac{\gamma }{4 k^3 \tau}-\frac{\gamma^3 \tau^3}{4 k^3}+\mathcal{O}\left(\frac{1}{k^5}\right)\,. 
\ee
In this case, and since we are directly looking at the energy density, we need to find coincidence with the adiabatic expansion up to and including order $\mathcal{O}(k^{-3})$.\footnote{Note that,  $\langle \rho \rangle\propto\int dk \, k^2 \rho_k(\tau)$, and therefore, a term going as $\sim k^{-3}$ generates a logarithmic divergence.} We directly see that, for $\Theta_k=0$, there is one term in \eqref{eq:dirac-largek}, $\gamma/(2k^3 \tau)$,  that does not coincide with the adiabatic expansion, generating a logarithmic divergence after renormalization. On the contrary, for $\Theta_k\sim-\frac{\gamma}{4 k^2}$, and using the large $k$ expansion of the trigonometric functions, we see that the part proportional to $\sin(2 \Theta_k)$ contributes to the large $k$ expansion in such a way that we recover the adiabatic expansion \eqref{eq:ad-largek}, making the energy density UV regular. Furthermore, in this case we find also cancellations for the oscillatory terms. We have also performed a similar analysis with the two-point function and the pressure, finding analogous results. We have found that, to ensure the convergence of these three quantities, the trigonometric phase should behave, at large $k$, as $\Theta_k \sim -\frac{\gamma}{4 k^2}+\mathcal{O}(k^{-n}), \ n >2$.

%S.P.~is supported by the Leverhulme Trust, Grant No.~RPG-2021-299.

%\section{antiguo: Review of vacuum choices in FLRW}We can show the emerging difficulties that appear in defining the vacuum at $\tau \to 0$
%\subsection{de Sitter vacuum $a(t)=e^{H t}$}
%Defined in the adiabatic regime and renders a Hadamard state with an exponential decay for particle creations.

%\subsection{radiation-dominated vacuum $a(\tau) \propto \tau$ }
%We describe the vacuum that minimizes $\rho_k$ at $\tau=0$, and recovers the conformal symmetry.

%\subsection{Case $a(t) \propto t $ }
%Limiting case where  we recover the massless positive frequency solution at $t \to 0$. Renders a Hadamard state, thermal particle creation at late times.

%\subsection{Case $a(t) \propto t^{1/3} $ }
%Maybe show this case too\\

%We can mention that if we want to obtain states which are more renormalizable we can cosntruct the corresponding states of low energy as we will illustrate in the following section for a radiation-dominated spacetime

%\cite{ex1,ex2,ex3}

\section*{References}
\bibliography{bibliography}
%\bibliography{iopart-num}
\end{document}